\documentclass[]{elsart}

\usepackage{amssymb,graphicx,subfigure,url, amsmath, rotating}
\usepackage{wasysym}

\newtheorem{proposition}{Finding}

\pdfoutput=1

\begin{document}

\begin{frontmatter}



\title{Steering plasmodium with light: \\ Dynamical programming of Physarum machine}

\author{Andrew Adamatzky}

\address{University of the West of England Bristol BS16 1QY United Kingdom\\ andrew.adamatzky@uwe.ac.uk}

\begin{abstract}
A plasmodium of \emph{Physarum polycephalum} is a very large cell visible by unaided eye. The plasmodium is
capable for distributed sensing, parallel information processing, and decentralized optimization. It is an 
ideal substrate for future and emerging bio-computing devices. We study space-time dynamics of plasmodium 
reactiom to localised illumination, and provide analogies between propagating plasmodium and travelling wave-fragments 
in excitable media.  We show how plasmodium-based computing devices can be precisely controlled and shaped 
by planar domains of illumination.
\end{abstract}


\end{frontmatter}

\section{Introduction}
\label{intro}

At one phase of its life-cycle \emph{Physarum polycephalum}\footnote{Order \emph{Physarales}, subclass \emph{Myxogastromycetidae}, class \emph{Myxomecetes}} lives as a huge single cell with many diploid nuclei. Such a life-form is called plasmodium. 
The plasmodium feeds on bacteria, spores and other microbial creatures. When foraging for its food the plasmodium propagates towards sources of food particles, surrounds them, secretes enzymes and digests the food.  Typically the plasmodium forms a congregation of protoplasm in a food source it occupies. When several sources of nutrients are scattered in the plasmodium's range, the plasmodium forms a network of protoplasmic tubes connecting the masses of protoplasm at the food sources. 

Nakagaki \emph{et al}~\cite{nakagaki_2000,nakagaki_2001,nakagaki_2001a} showed that the topology of the  plasmodium's protoplasmic 
network optimizes the plasmodium's harvesting on distributed sources of nutrients and makes more efficient flow
and transport of intra-cellular components. The plasmodium is considered as a parallel computing substrate 
complementary~\cite{adamatzky_naturewissenschaften_2007} to existing massive-parallel reaction-diffusion 
chemical processors~\cite{adamatzky_2005}. The plasmodium functions as a parallel amorphous computer with parallel inputs. 
Such a parallel biological computer takes data in a form of spatially distributed sources of nutriets, and represents
results of computation by configuration of its entire body and protoplasmic veins. The plasmodium is capable for a
approximation of shortest path, computation of planar proximity graphs and plane tessellations, primitive memory and decision-making.

In~\cite{adamatzky_ppl_2007} we shown that the plasmodium of \emph{Physarum polycephalum} is a general 
purpose computing machine. This is because plasmodium implements the Kolmogorov-Uspensky (KUM) machine~\cite{kolmogorov_1953,uspensky_1992} in its foraging behaviour. The KUM is a mathematical machine 
in which the storage structure is an irregular graph. The KUM is a forerunner and direct `ancestor' of Knuth's linking automata~\cite{knuth_1968}, Tarjan's reference machine~\cite{tarjan_1977}, and  Sch\"{o}nhage's storage modification machines~\cite{schonhage_1973,schonhage_1980}. The storage modification machines are basic architectures for random access machines, which are the basic architecture of modern-day computers. The plasmodium-based implementation of KUM~\cite{adamatzky_ppl_2007} is a first-ever biological prototype of a general purpose computer.

The key component of the KUM is an active zone~\cite{kolmogorov_1953,uspensky_1992}, which may be seen as a computational
equivalent to the head in a Turing machine. Physical control of the active zone is of utmost importance because it determines
functionality of the biological storage modification machine. In present paper we experimentally demonstrate that 
propagation of KUM's active zone can be tuned by localized domains of illumination.

The plasmodium of \emph{Physarum} exhibits negative phototaxis. General understanding of the plasmodium response to light is 
that the plasmodium moves away from light when it can and switches to another phase of its life cycle 
or undergoes phragmentation when it could not escape light. If plasmodium, particularly the 
starving one~\cite{guttes_1961}, is subjected to high intensity of light the plasmodium turns 
into a sporulation phase~\cite{sauer_1969}. There are evidences that phytochromes are involed in the light-induced sporulation~\cite{starostzik_1995}  and sporulation morphogene is transferred by protoplasmic streams 
to all parts of the plasmodium~\cite{Hildebrandt_1986}. 

Photofragmentation is another physiological response to strong and unavoidable illumination. When plasmodium is illuminated
by ultraviolet or blue monochromatic light, in hostile environment of laboratory conditions, it breaks into many equally 
sized fragments (each fragment contains around eight nuclei)~\cite{kakiuchi_2001}. The fragmentation is transient and 
after some time the fragments merge back into a fully functional plasmodium.    

In the paper we focus on photomovement, the less drastic response to illumination than two scenarios mentioned above.
Plasmodium of \emph{Physarum polycephalum} exhibits most pronounced negative phototaxis to blue and white light~\cite{bailczyk_1979,schreckenbach}. The illumination increase causes changes in the plasmodium oscillatory 
activity, degree of changes is proportional to a distance from the light source~\cite{Wohlfarth-Bottermann_1981,block_1981}.
Exact mechanisms of the response to light is yet unknown. There is however a few phenomena uncovered in experiments.
First is presence of phytochrome-like pigments~\cite{kakiuchi_2001}, which might be primary receptors of illumination. 
The light-response of the pigments triggers a chain of biochemical process~\cite{schreckenbach_1980}. These processes
include increase in activity of isomerase enzymes~\cite{starona_1992}, changes in mytochondrial respiration~\cite{korohoda_1983},
and spatially distributed oscillations in ATP concentrations~\cite{ueda_1986}. 

Nakagaki et al~\cite{nakagaki_yamada_1999,nakagaki_iima_2007} undertook first ever experiments on shaping plasmodium 
behavior with illumination.  They discovered that protoplasm streaming oscillations of  plasmodium can be tuned by,
or relatively synchronized with, periodic illumination~\cite{nakagaki_yamada_1999}. They also demonstrated that plasmodium optimizes its protoplasmic network structure in the field with heterogeneous illumination~\cite{nakagaki_iima_2007}: thickness of protoplasmic tubes in illuminated areas  are less then thickness of tubes in shaded areas~\cite{nakagaki_iima_2007}.
These indicate that illumination gradients could be a convenient tool to input instructions to different parts of \emph{Physarum} machines in parallel. Several basic questions need to be answered. What exactly a pseudopodium or a migrating \emph{Physarum}
do when approach an illuminated domain? How light can be used to program plasmodium movements? What types of plasmodium reflections can be implemented using light-mirrors? Can we shape the plasmodium's network structure by heterogeneous illumination?  In the paper we present experimental and theoretical findings we obtained while trying to answer the questions.

\section{Methods}

The plasmodia of \emph{Physarum polycephalum}\footnote{Culture of \emph{Physarum polycephalum} was kindly supplied by 
Dr. Soichiro Tsuda} were cultured on wet paper towels, fed with oat flakes, and moistened regularly. We subcultured the plasmodium every 5-7 days. 

\begin{figure}[tb]
\centering
\subfigure[]{\includegraphics[width=0.9\textwidth]{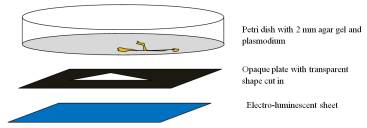}}
\subfigure[]{
\includegraphics[width=0.3\textwidth]{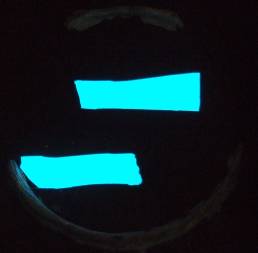}
\includegraphics[width=0.3\textwidth]{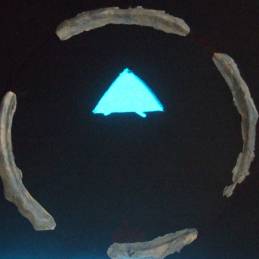}
\includegraphics[width=0.3\textwidth]{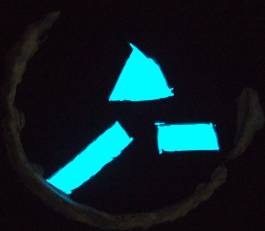}
}
\caption{Experimental setup: (a)~scheme, (b)~photographs of illuminating domains, geometrical shapes.}
\label{experimentalsetup}
\end{figure}

Experiments on controlling the plasmodium with light were undertaken in standard Petri dishes, 9~cm in diameter. A substrate was 2\% agar gel. Light-obstacles, or illumination domains, were implemented using electro-luminescent sheets~\footnote{Manufacturer Seikosha, supplier RS Components Ltd Birchington Road, Corby, Northants, NN17 9RS, UK}. Based on previous works in negative phototaxis  of \emph{Physarum} we have choosen blue illuminating sheets. The nominal sheets brightness (at 110 V, 400 Hz AC) is 73~cd/m$^2$ (blue). The electro-luminiscent sheets did not produre heat, therefore can be positioned in close contact with growing substrate. Masks were prepared from black plastic: rectangles and triangles were cut  in the plastic. When a mask is placed on top of the electro-luminiscent sheet, the light is passed only throught the cuts (Fig.~\ref{experimentalsetup})\footnote{Pictures quality is dramatically reduced. See version published in the Journal for high-resolution pictures.}.

The experiments were conducted in a room with diffusive light of 3-5~cd/m, 22$^o$C temperature. In each experiment an oat flake colonized by the plasmodium was placed on one side
of a Petri dish, and few oat flakes without plasmodium at the opposite side of the Petri dish (to shape directional propagation 
of the plasmodium). Petri dishes with plasmodium were periodically scanned on a standard HP scanner. The only editing done to scanned images is colour enhancement: increase of saturation and contrast.

When plasmodium is cultivated on a non-nutrient substrate a profile of its pseudopodium's propagating tip, and particularly the profile of the plasmodium propagating as a whole, is isomorphic to shapes of and behaves as wave-fragments in sub-excitable reaction-diffusion chemical media, see details in~\cite{adamatzky_delacycostello_shirakawa_2008,adamatzky-bz-trees}. Therefore we simulated propagation of the plasmodium with two-variable Oregonator equation~\cite{field_noyes_1974,tyson_fife}, used originally to  simulate light-sensitive Belousov-Zhabotinsky reaction with applied gradients of illumination~\cite{beato_engel,krug}:

$$\frac{\partial u}{\partial t} = \frac{1}{\epsilon} (u - u^2 - (f v + \phi)\frac{u-q}{u+q}) + D_u \nabla^2 u$$
$$\frac{\partial v}{\partial t} = u - v .$$

In terms of Belousov-Zhabotisnky reaction the variables $u$ and $v$ represent local concentrations of bromous acid HBrO$_2$
and the oxidized form of the catalyst ruthenium Ru(III), $\epsilon$ sets up a
ratio of time scale of variables $u$ and $v$, $q$ is a scaling parameter depending
on reaction rates, $f$ is a stoichiometric coefficient, $\phi$ is a light-induced
bromide production rate proportional to intensity of illumination. The $\phi$ is an excitability parameter. Moderate intensity of light will facilitate excitation process, higher intensity will produce excessive quantities of bromide which suppresses the reaction. There is no diffusion term for $v$ because we assume the catalyst is immobilized.

To integrate the system we used Euler method with five-node Laplasian operator, null boundary conditions.
time step $\Delta t=5\cdot10^{-3}$ and grid point spacing $\Delta x = 0.25$, with the 
following parameters:
$\phi=\phi_0 - \alpha/2$, 
$A=0.0011109$, 
$\phi_0=0.0804$, 
$\epsilon=0.03$, 
$f=1.4$, 
$q=0.022$.

Illuminated domain (light obstacle) $S$ of the experimental space was simulated by higher values of the parameter
$\phi$: if given point belongs to $S$ then $\phi=0.085$ (the medium inside the light obstacle becomes non-excitable) otherwise 
$\phi=0.0804$. For such set of parameters the model represents a sub-excitable medium (at the edge between of non-excitability and full excitability). The sub-excitable media exhibits self-localized excitations. A local disturbance leads to formation of traveling excitation wave-fragments which preserve their shape for finite period of time and then either expand or collapse. The traveling wave-fragments imitate propagating plasmodium.

\section{Trees and waves}

There are two distinct forms of the plasmodium: a protoplasmic tree and a traveling localizations. 
The protoplasmic tree is formed when plasmodium forages the space by sprouting pseudopodia in various directions. 
The pseudopodia remain connected to original location (and still main ``body'') of the plasmodium by protoplasmic tubes. 
In certain situations (exact characteristics of such situations might be a subject of further studies) the plasmodium 
leaves its original location and propagates as a whole on the substrate. When the plasmodium propagates as a whole (migrating plasmodium) 
it has shape
of a wave-fragment traveling in the a sub-excitable medium. The boundary between two morphologies may be fuzzy, and very often 
we can observe wave-fragment like shapes of pseudopodia, and gradual transitions between tree-like and wave-fragment like morphology. 

\begin{figure}[tb]
\centering
\subfigure[$t=0$~h]{\includegraphics[width=0.3\textwidth]{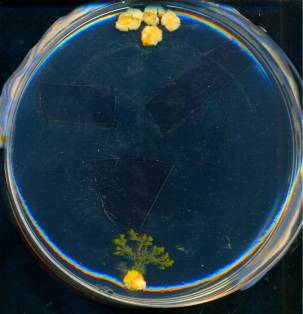}}
\subfigure[$t=3$~h]{\includegraphics[width=0.3\textwidth]{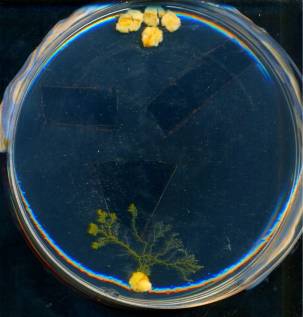}}
\subfigure[$t=6$~h]{\includegraphics[width=0.3\textwidth]{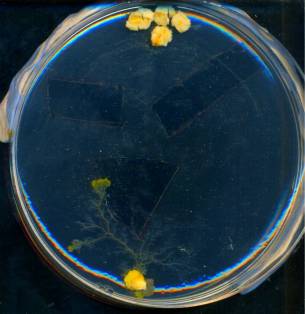}}
\subfigure[$t=11$~h]{\includegraphics[width=0.3\textwidth]{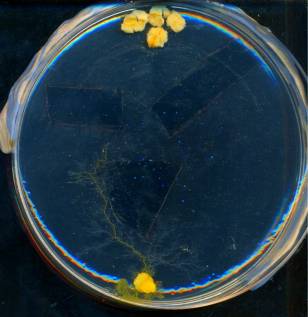}}
\subfigure[$t=20$~h]{\includegraphics[width=0.3\textwidth]{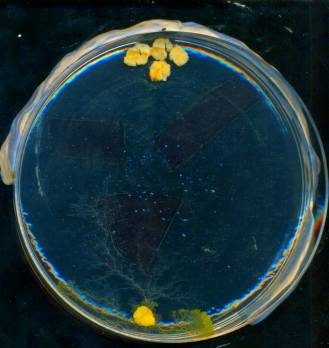}}
\subfigure[$t=29$~h]{\includegraphics[width=0.3\textwidth]{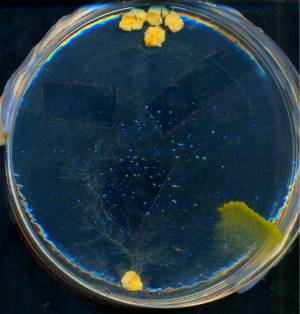}}
\subfigure[$t=33$~h]{\includegraphics[width=0.3\textwidth]{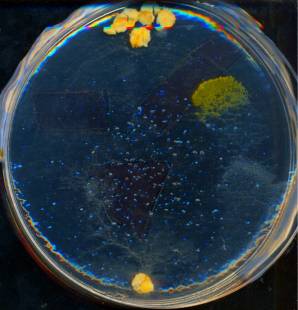}}
\subfigure[$t=39$~h]{\includegraphics[width=0.3\textwidth]{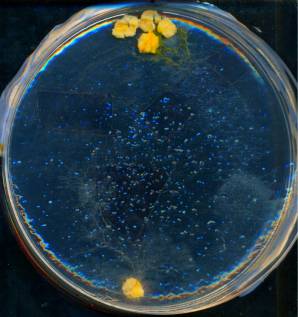}}
\caption{Transformation of plasmodium from protoplasmic tree to travelling localization (migrating plasmodium). }
\label{tree2wave}
\end{figure}

An example of the ``tree to wave-fragment'' transformation is provided in Fig.~\ref{tree2wave}. An oat flake colonized 
by plasmodium is placed in south part of a Petri dish (Fig.~\ref{tree2wave}a). The plasmodium sprouts several pseudopodia 
exploring the space (Fig.~\ref{tree2wave}a). The protoplasmic branches die out when encounter illuminated domains
(Fig.~\ref{tree2wave}bc). A group of branching pseudopodia tries to find a way around the triangular light obstacle
(Fig.~\ref{tree2wave}c) but eventually abandons the attempt (Fig.~\ref{tree2wave}d). 

By that time bacteria on the original oat flakes are exhausted and the plasmodium switches to its migration phase (Fig.~\ref{tree2wave}d). The plasmodium abandons its original oat flake and  starts propagating {\emph as a whole} 
along the eastern wall of the Petri dish (Fig.~\ref{tree2wave}ef). A typical wave-fragment of the propagating 
plasmodium is formed (Fig.~\ref{tree2wave}g) which heads toward the source of nutrients (a group of oat flakes 
in the northern part of Petri dish). Eventually the plasmodium reaches new source of nutrients (Fig.~\ref{tree2wave}h). 

When the plasmodium propagates as {\emph a whole} it looks like and behaves like a wave-fragment in sub-excitable 
medium~\cite{adamatzky_delacycostello_shirakawa_2008,adamatzky-bz-trees}. Moreover, migrating plasmodium is more sensitive to 
potential environment threats than just propagating pseudopodia. This is because a protoplasm in pseudopodia can always be retracted back to main body of plasmodium, thus pseudopodia can ``take risks''. The migrating plasmodium is more vulnerable, because
any ``mis-calculation'' in choosing its migration route may lead to disaster.

\section{Diverting plasmodium}

In an ideal situation the plasmodium propagates as wave-fragment and --- unless encounters an obstacle on the way ---
keeps its shape and velocity vector conserved. If a proximal part of the plasmodium wave comes upon a highly
illuminated domain, frequency of protoplasm oscillations in this domain increases. Due to a difference in the protoplasm
oscillation frequency the plasmodium wave slightly turns to the side with less oscillating protoplasm.

\begin{proposition}
One can steer propagating plasmodium using light obstacles. 
\end{proposition}

\begin{figure}[tb]
\centering
\subfigure[$t=0$~h]{\includegraphics[width=0.3\textwidth]{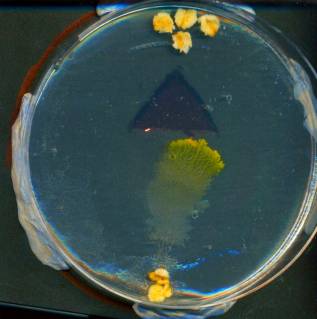}}
\subfigure[$t=$1~h]{\includegraphics[width=0.3\textwidth]{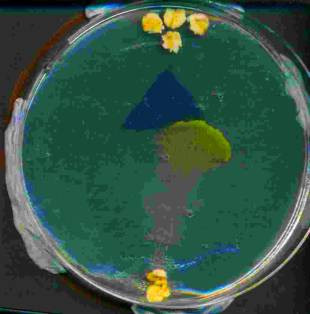}}
\subfigure[$t=$5~h]{\includegraphics[width=0.3\textwidth]{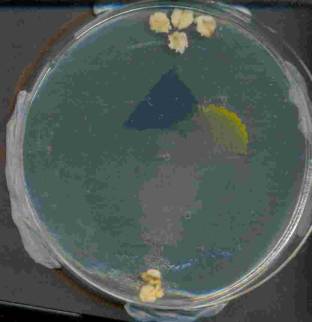}}
\subfigure[$t=$11~h]{\includegraphics[width=0.3\textwidth]{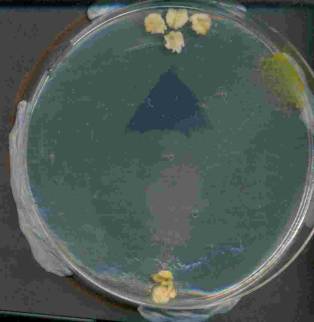}}
\subfigure[$t=$0~h]{\includegraphics[width=0.3\textwidth]{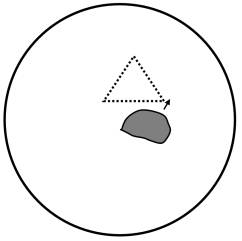}}
\subfigure[$t=$1~h]{\includegraphics[width=0.3\textwidth]{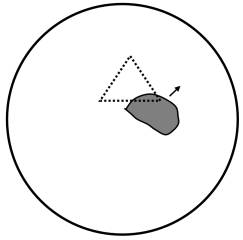}}
\subfigure[$t=$5~h]{\includegraphics[width=0.3\textwidth]{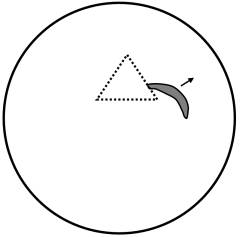}}
\subfigure[$t=$11~h]{\includegraphics[width=0.3\textwidth]{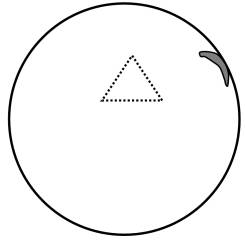}}
\caption{Deflection of a plasmodium wave by light triangle. (a)--(d)~photographs of the plasmodium, 
(e)--(h)~schemes of the propagating plasmodium waves.}
\label{deflectiontriangle}
\end{figure}

A deflection of the plasmodium wave by light triangle is demonstrated in Fig.~\ref{deflectiontriangle}. The plasmodium wave propagates North-North-East (Fig.~\ref{deflectiontriangle}ae). The plasmodium hits a light triangle with its 
western side (Fig.~\ref{deflectiontriangle}bf). The light increases frequency of oscillations in the 
illuminated part of the plasmodium. The plasmodium wave turns eastward (Fig.~\ref{deflectiontriangle}cg) and travels in the new direction until hits the dish's wall (Fig.~\ref{deflectiontriangle}cdh).

\begin{figure}[tb]
\centering
\subfigure[$t=0$~h]{\includegraphics[width=0.3\textwidth]{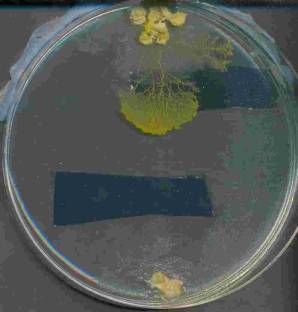}}
\subfigure[$t=7$~h]{\includegraphics[width=0.3\textwidth]{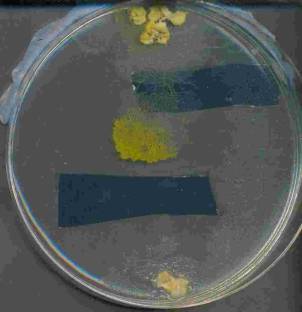}} \\
\subfigure[$t=0$~h]{\includegraphics[width=0.3\textwidth]{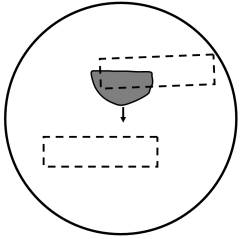}}
\subfigure[$t=7$~h]{\includegraphics[width=0.3\textwidth]{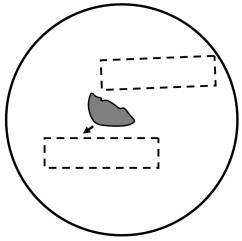}}
\caption{Deflection of plasmodium wave by light rectangle. (a)--(d)~photographs of the plasmodium, 
(e)--(h)~schemes of the propagating plasmodium waves.}
\label{deflectionrectangle2}
\end{figure}

The plasmodium may not reflect from light obstacle immediately. After encountering even highly illuminated area the plasmodium wave carries on traveling in the original direction and only after passing obstacle starts diverting. Such example of delayed reaction to illumination is shown in Fig.~\ref{deflectionrectangle2}. Plasmodium wave, originated from the plasmodium colonizing a 
group of flakes in northern part of the dish, travels South (Fig.~\ref{deflectionrectangle2}ac). The wave passes through rectangular illuminated area without any immediate reaction (Fig.~\ref{deflectionrectangle2}a). The response occurs several hours later. The plasmodium steers South-West (Fig.~\ref{deflectionrectangle2}bd). Note that the plasmodium is \emph{not} diverted by second (lying southward) light rectangle, because when diversion happens the plasmodium is far from the second light rectangle. 

\begin{figure}[tb]
\centering
\subfigure[$t=0$~h]{\includegraphics[width=0.3\textwidth]{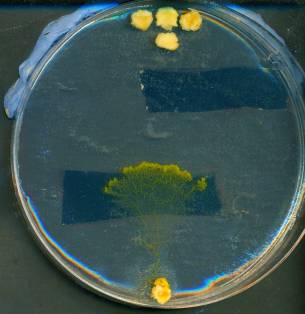}}
\subfigure[$t=12$~h]{\includegraphics[width=0.3\textwidth]{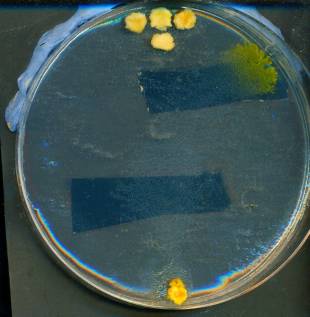}} \\
\subfigure[$t=0$~h]{\includegraphics[width=0.3\textwidth]{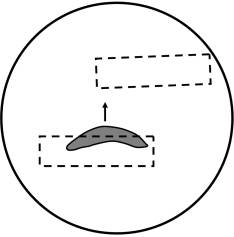}}
\subfigure[$t=12$~h]{\includegraphics[width=0.3\textwidth]{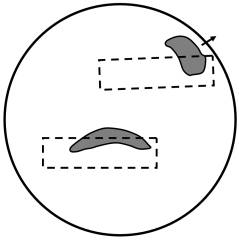}} \\
\caption{Deflection of plasmodium wave by light rectangle. (a)--(d)~photographs of the plasmodium, 
(e)--(h)~schemes of the propagating plasmodium waves.}
\label{deflectionrectangle1}
\end{figure}

Another example of the steering by light obstacles is shown in Fig.~\ref{deflectionrectangle1}. The plasmodium wave, heading North, enters the illuminated rectangular area (Fig.~\ref{deflectionrectangle1}ac). The wave is slightly displayed towards eastern side 
of the light rectangle. Due to differences in light-induced oscillation of protoplasm the plasmodium turns North-East and continues
traveling in this direction till collides with a wall of the Petri dish (Fig.~\ref{deflectionrectangle1}bd).

\begin{proposition}
Combining light obstacles and sources of nutrients (chemo-attractants) one can implement precise control of plasmodium waves.
\end{proposition}

\begin{figure}[tb]
\centering
\subfigure[$t=0$~h ]{\includegraphics[width=0.3\textwidth]{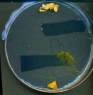}}
\subfigure[$t=12$~h]{\includegraphics[width=0.3\textwidth]{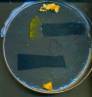}}
\subfigure[$t=16$~h]{\includegraphics[width=0.3\textwidth]{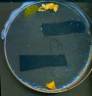}}
\subfigure[$t=0$~h ]{\includegraphics[width=0.3\textwidth]{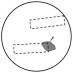}}
\subfigure[$t=12$~h]{\includegraphics[width=0.3\textwidth]{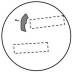}}
\subfigure[$t=16$~h]{\includegraphics[width=0.3\textwidth]{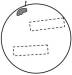}}
\caption{Steering plasmodium with combination of light obstacles and attracting field.}
\label{deflectionrectangle}
\end{figure}

The plasmodium is attracted to oat flakes populated with bacteria: we experimentally demonstrated that the plasmodium senses and reacts to sources of nutrients placed as far as 3-4 cm away of the plasmodium. The plasmodium `ascents' along gradients of 
the attractants till it reaches sources of nutrients. By placing light obstacles in the attractive field we can tune and shape 
trajectory of the plasmodium motion. For example, the plasmodium shown in Fig.~\ref{deflectionrectangle} is reflected eastward
by first illuminated rectangle it encounters (Fig.~\ref{deflectionrectangle}ad). Later the plasmodium turns North being attracted to oat flakes in Northern part of the Petri dish. The plasmodium is directed westward by the second illuminated 
rectangle (Fig.~\ref{deflectionrectangle}be). However it does not continue West but turns North again to reach the sources of nutrients (Fig.~\ref{deflectionrectangle}cf).

\section{Multiplying plasmodium waves}

\begin{proposition}
A propagating plasmodium wave or a pseudopodium can be split by suitably shaped domain of illumination.
\end{proposition}

\begin{figure}[tb]
\centering
\subfigure[0~h]{\includegraphics[width=0.3\textwidth]{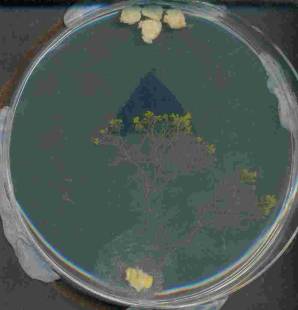}}
\subfigure[6~h]{\includegraphics[width=0.3\textwidth]{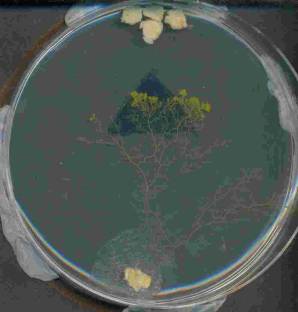}}
\subfigure[11~h]{\includegraphics[width=0.3\textwidth]{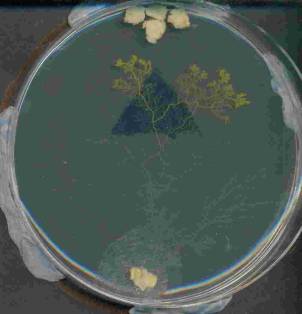}}
\subfigure[0~h]{\includegraphics[width=0.3\textwidth]{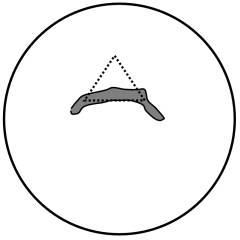}}
\subfigure[6~h]{\includegraphics[width=0.3\textwidth]{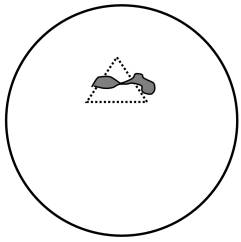}}
\subfigure[11~h]{\includegraphics[width=0.3\textwidth]{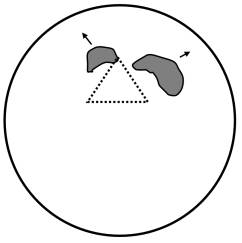}}
\caption{Splitting plasmodium wave with illuminated triangle.}
\label{splittingtriangle}
\end{figure}

In some situation propagating plasmodium hits a light obstacle which is small enough to divert the whole 
plasmodium wave. If parts of the plasmodium wave remain outside the illuminated shape, these parts continue 
travel as ``independent'' plasmodium waves. Thus the plasmodium wave will be split into two waves. An example
is provided in Fig.~\ref{splittingtriangle}.  In this particular experiment the plasmodium wave is comprised of several 
pseudopodia moving together in a close group. On its way toward source of nutrients the wave runs across illuminated 
triangle (Fig.~\ref{splittingtriangle}ad). The pseudopodia try to steer away from the source of light and move toward
western and eastern sides of the illuminated triangle (Fig.~\ref{splittingtriangle}be). Eventually two separate groups of
pseudopodia are formed, one group travels North-West, second group moves North-East (Fig.~\ref{splittingtriangle}be).

\section{Optimization of foraging in presence of illumination-obstacles}

\begin{proposition}
When \emph{Physarum} scouts a space with pseudopodia it abandons the pseudopodia which encounter illuminated domains and 
sprouts new pseudopodia in the less illuminated areas and towards the sources of nutrients.
\end{proposition}

Nakagaki {\emph et al}~\cite{nakagaki_iima_2007} already demonstrated that the plasmodium optimizes its protoplasmic network in 
presence of light, namely the plasmodium reduces number and size of protoplasmic tubes exposed to light. In present paper 
we do not focus on optimization and do not provide statistical evidences but rather few morphological examples of how 
growing protoplasmic network is shaped by illuminated domains. 

\begin{figure}[tb]
\centering
\subfigure[$t=0$~h]{\includegraphics[width=0.3\textwidth]{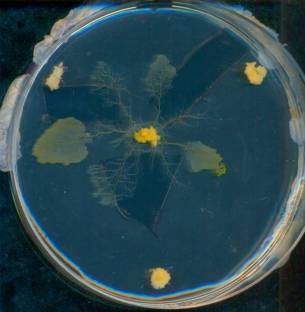}}
\subfigure[$t=9$~h]{\includegraphics[width=0.3\textwidth]{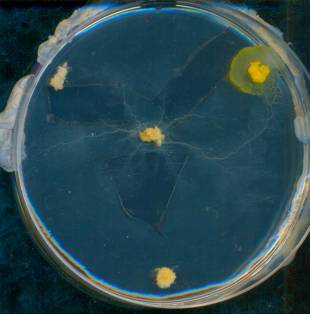}}\\
\subfigure[$t=0$~h]{\includegraphics[width=0.3\textwidth]{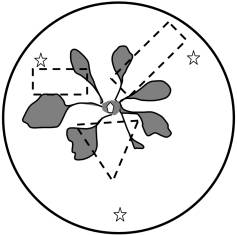}}
\subfigure[$t=9$~h]{\includegraphics[width=0.3\textwidth]{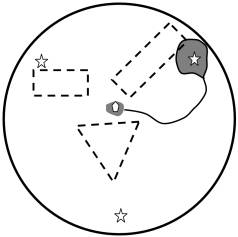}}\\
\caption{Plasmodium scouting in presence of illumination obstacles. (a)--(b)~experimental, (c)--(d)~schemes of propagation.
Oat flakes are shown by stars. }
\label{scountingfromcentre2}
\end{figure}

Most typical example is shown in Fig.~\ref{scountingfromcentre2}. Initially an oat flake colonized by the plasmodium is 
placed in a central part of the Petri dish (Fig.~\ref{scountingfromcentre2}ac). Three intact oat flakes are placed near the Petri dish wall so they form vertexes of a triangle centered in the initial position of the plasmodium. The plasmodium sprouts
several pseudopodia heading towards the oat flakes. Most of the pseudopodia encounter illuminated areas and seize propagation. 
The pseudopodium propagating East-South-East manages to avoid illuminated areas and eventually reaches North-East oat flake (Fig.~\ref{scountingfromcentre2}bd).  

\begin{figure}[tb]
\centering
\subfigure[0~h]{\includegraphics[width=0.3\textwidth]{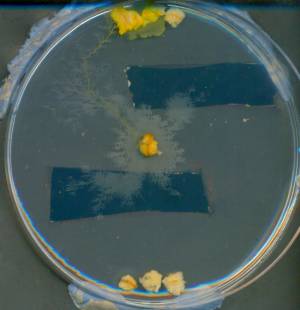}}
\subfigure[0~h]{\includegraphics[width=0.3\textwidth]{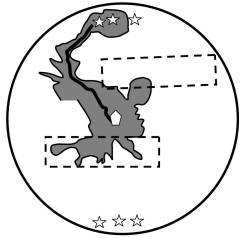}} \\
\caption{Plasmodium scouting in presence of illumination obstacles. (a)~snapshot of experiments, 
(b)~scheme of propagation. Oat flakes are shown by stars.}
\label{scountingfromcentre}
\end{figure}

There are evidences that the plasmodium may ``evaluate'' size of the light obstacles by sprouting pseudopodia 
in several directions, integrating information about size of obstacles (possible via protoplasm oscillation frequencies) and then
making a decision about where to sprout further pseudopodia. An example of such decision making is provided in 
Fig.~\ref{scountingfromcentre}. In the experiment reported the plasmodium was placed in the center of a Petri dish, and additional 
oat flakes were placed in North and South poles of the dish. The plasmodium sprouts several pseudopodia to explore the space. 
Some of the pseudopodia encounter illuminated areas. The illuminated area south of the plasmodium is larger than the illuminated area north of the plasmodium. Therefore the plasmodium does not sprout any more pseudopodia southward but produces a protoplasmic
branch growing North. This pseudopodium reaches the south source of nutrients (Fig.~\ref{scountingfromcentre}). 

\begin{figure}[tb]
\centering
\subfigure[]{\includegraphics[width=0.3\textwidth]{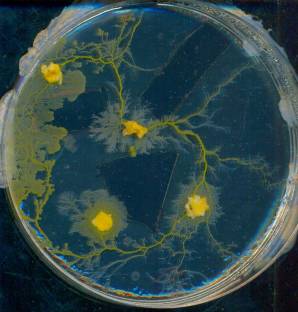}}
\caption{Protoplasmic network shaped by light obstacles.}
\label{network}
\end{figure}

Such mechanism of scouting allow plasmodium to pre-shape its foraging network. When eventually plasmodium spans all 
sources of nutrients with its protoplasmic network major tubes of the network become positioned in the non-illuminated areas (Fig.~\ref{network}).

\section{Simulating plasmodium in two-variable Oregonator}

\begin{figure}[tb]
\centering
\subfigure[$t=200$]{\includegraphics[width=0.2\textwidth]{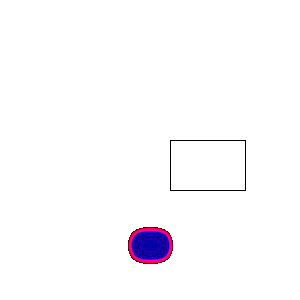}}
\subfigure[$t=400$]{\includegraphics[width=0.2\textwidth]{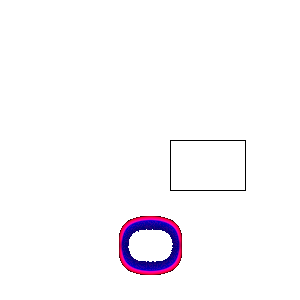}}
\subfigure[$t=600$]{\includegraphics[width=0.2\textwidth]{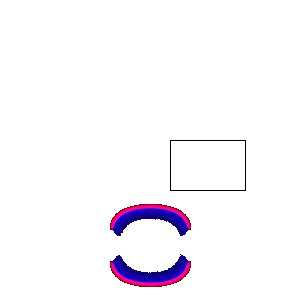}}
%\subfigure[$t=800$]{\includegraphics[width=0.2\textwidth]{figs/rectangle-deflect/bz-0800}}
\subfigure[$t=1000$]{\includegraphics[width=0.2\textwidth]{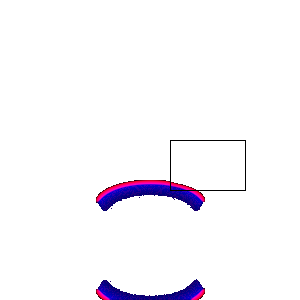}}
\subfigure[$t=1200$]{\includegraphics[width=0.2\textwidth]{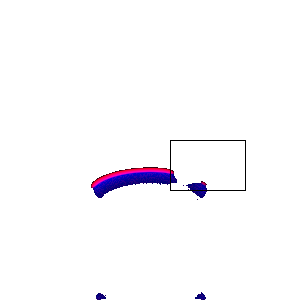}}
\subfigure[$t=1400$]{\includegraphics[width=0.2\textwidth]{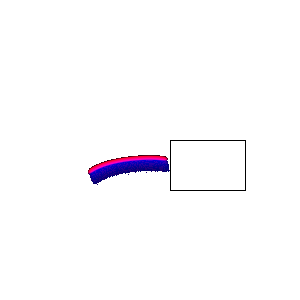}}
\subfigure[$t=1600$]{\includegraphics[width=0.2\textwidth]{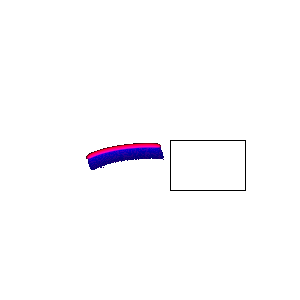}}
\subfigure[$t=1800$]{\includegraphics[width=0.2\textwidth]{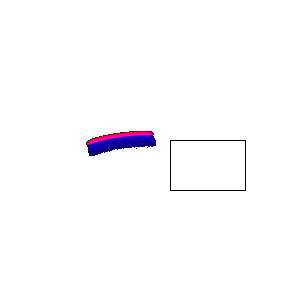}}
\caption{Simulation of plasmodium wave diverted by recnangular illuminated domain. Red and blue components of each pixel's color were defined as follows.  If $u>0.1$ then red value $255\cdot u$, if $v>0.1$ then blue value $600\cdot v$, otherwise the background white color used. Time shows steps of numerical integration.}
\label{rectangledeflectOregonator}
\end{figure}

Striking similarity in behaviour of the plasmodium and wave-fragments in sub-excitable media has been 
found in our previous papers~\cite{adamatzky_delacycostello_shirakawa_2008,adamatzky-bz-trees}. Oregonator based model 
of the plasmodium waves impeccably matches our experimental results. An example of simulation is provided in 
Fig.~\ref{rectangledeflectOregonator}. 

At the beginning of experiments a piece of plasmodium is simulated by exciting the medium 
with ellipse-shaped stimulus (Fig.~\ref{rectangledeflectOregonator}a): the medium is perturbed by an initial excitation, 
when $15\times15$ sites are assigned $u=1.0$ each.
The perturbation generates an ellipse-shaped excitation wave. Due to sub-excitability of the simulated medium, the
initially circular wave front (Fig.~\ref{rectangledeflectOregonator}b) breaks into two wave fragments. One wave travels North, another wave South (Fig.~\ref{rectangledeflectOregonator}c--e). We discard South travelling wave. When the Eastern part of the North travelling wave enters illumination area it becomes inhibited and gradually disappears (Fig.~\ref{rectangledeflectOregonator}ef). The remaining part of the wave continues its travel as an individual wave, propagating North-North-West (Fig.~\ref{rectangledeflectOregonator}g--i). 

\begin{figure}[tb]
\centering
\subfigure[$t=100$]{\includegraphics[width=0.25\textwidth]{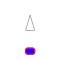}}
\subfigure[$t=400$]{\includegraphics[width=0.25\textwidth]{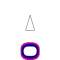}}
\subfigure[$t=600$]{\includegraphics[width=0.25\textwidth]{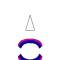}}
\subfigure[$t=800$]{\includegraphics[width=0.25\textwidth]{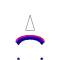}}
\subfigure[$t=1000$]{\includegraphics[width=0.25\textwidth]{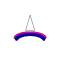}}
\subfigure[$t=1200$]{\includegraphics[width=0.25\textwidth]{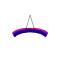}}
\subfigure[$t=1400$]{\includegraphics[width=0.25\textwidth]{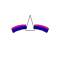}}
\subfigure[$t=1600$]{\includegraphics[width=0.25\textwidth]{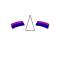}}
\subfigure[$t=1800$]{\includegraphics[width=0.25\textwidth]{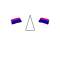}}
\caption{Simulation of plasmodium wave splitting  by illuminated triangle. 
Red and blue components of each pixel's color were defined as follows.  If $u>0.1$ then red value $255\cdot u$, 
if $v>0.1$ then blue value $600\cdot v$, otherwise the background white color used. Time shows steps of 
numerical integration.}
\label{trianglesplitOregonator}
\end{figure}

Simulation results for wave splitting are shown in Fig.~\ref{trianglesplitOregonator}. 
An ellipse-shaped zone of initial excitation (Fig.~\ref{trianglesplitOregonator}a) produces two wave-fragments 
travelling North and South (Fig.~\ref{trianglesplitOregonator}bc). We are concerneted with the fragment propagating
North (Fig.~\ref{trianglesplitOregonator}de). When the wave-fragment reaches illuminated triangle the part of the wave inside
the triangle extinguishes due to excitability inhibited by light (Fig.~\ref{trianglesplitOregonator}fg). In the result of the 
inhibition two separate wave-fragments are formed. One wave-fragment travels South-West another travels South-East (Fig.~\ref{trianglesplitOregonator}hi).

\section{Routing signals in Physarum machine}

Kolmogorov-Uspensky Machine (KUM) is defined on a labeled undirected graph (storage structure) with bounded degrees of nodes and bounded number of labels~\cite{kolmogorov_1953,uspensky_1992}. KUM executes the following operations on its storage structure:
select an active node in the storage graph; specify local active zone, i.e. the node's neighborhood; 
modify the active zone by adding a new node with the pair of edges, connecting the new node with the active node; delete a node with a pair of incident edges; add/delete the edge between the nodes. A program for KUM specifies how to replace the neighborhood of an active node (i.e. occupied by an active zone) with a new neighborhood, depending on the labels of edges connected to the active node and the labels of the nodes in proximity of the active node~\cite{blass_gurevich_2003}.

Physarum machine is a biological implementation of KUM, where a node of the storage structure is represented by a source of nutrients (e.g. an oat flake); an edge connecting two nodes is a protoplasmic tube linking two sources of nutrients corresponding 
to the nodes; and, an {\emph active zone} is domain of space (which may include food sources) occupied by a propagating pseudopodium. A migrating plasmodium also represents an active zone.

In Physarum machine the computation is implemented by active zone, or several active zones. To make the computation process
programmable one needs to find ways of sensible and purposeful manipulation with the active zones. In~\cite{adamatzky_jones_NC} 
we experimentally demonstrated how active zones can be manipulated by dynamical addition of nutrients. Programming with nutrients 
is not really efficient, because once the source of nutrients placed in the computing space, it irreversibly changes 
configuration of attracting fields. Light inputs allow us for on-line reconfiguration of obstacles and thus provide more 
opportunities for embedding complex programs in Physarum machines.

\begin{figure}[tb]
\centering
\subfigure[\sc{Erase}$(A)$]{\includegraphics[width=0.3\textwidth]{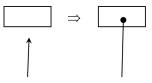}} \hspace{0.5cm}
\subfigure[\sc{Right}$(A)$]{\includegraphics[width=0.3\textwidth]{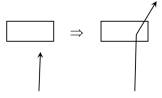}}\\
\subfigure[\sc{Left}$(A)$]{\includegraphics[width=0.3\textwidth]{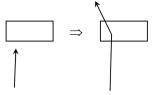}} \hspace{0.5cm}
\subfigure[\sc{Multiply}$(A)$]{\includegraphics[width=0.3\textwidth]{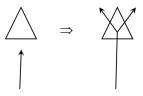}}
\caption{Scheme of routing operations in Physarym machine. Directions of active zone movement are shown by arrows. Illuminated domains are
rectangular and triangular shapes.}
\label{routing}
\end{figure}   

Basic operations of routing active zones are shown in Fig.~\ref{routing}.  

Operation {\sc Erase}$(A)$  removes, or permanently deactivates, active zone $A$ (Fig.~\ref{routing}a). The operation is implemented by placing large enough -- so there is no chance of the plasmodium wave to escape the shaded area --  domain of illumination 
in front of the traveling zone $A$. The plasmodium wave $A$ then gradually disappears.  The operation is experimentally exemplified 
in Fig~\ref{scountingfromcentre2}, where few active zones are canceled by rectangles of illumination. 

Operation {\sc Left}$(A)$ rotates velocity vector of zone $A$ on  angle $\alpha$, $-90 \leq \alpha < 0$; 
for {\sc Rigth}$(A)$ $0 < \alpha \leq 90$ (Fig.~\ref{routing}bc). Exact value of rotation angle $\alpha$ depends on many factors, including size of illumination domain, size of traveling zone $A$, humidity, overall illumination, and fitness of the plasmodium. Based on  few successful experiments on rotation of plasmodium waves, we could say that $\alpha = 45^o \pm 15^o$.

Operation {\sc Multiply}$(A)=\{ A_L, A_R \}$ (Fig.~\ref{routing}bc) splits traveling plasmodium wave $A$ into zones $A_L$ and $A_R$ deviated slightly left and right, comparing to original direction of $A$'s travel.

\section{Discussion}

In laboratory experiments with plasmodium of \emph{Physarum polycephalum} and in computer simulations 
we discovered that propagating plasmodium interacts with illuminated domains similarly to travelling wave-fragments 
in excitable media. We demonstrated that plasmodium waves can be diverted, annihilated and split 
by properly arranged shapes of illumination. The light-induced diversion of the plasmodium waves 
can be used as operations of signal and process routing in plasmodium-based implementations of 
general purpose storage machines~\cite{adamatzky_ppl_2007}. Future studies will concern with programming Physarum machines by combination of chemoattractants' gradients and illumination domains.

Usually quite high intensity of illumination is used in studies of \emph{Physarum} response~\cite{block_1981,nakagaki_iima_2007}. 
Thus in \cite{block_1981} reported threshold intensity for blue light is about 1,500~Lux. In our experiments the plasmodium 
was controlled by much less illuminated shapes, with maximum intensity of illumination of 50-70~Lux. This is why we did not 
observe strong photophobic reaction but rather gentle and often subtle changes in \emph{Physarum} behavior. The moderate 
intensity of illumination used in our experiments shows that we can achieve purposeful behavior of the plasmodium with light stimulation without causing any adverse reactions. 

Geometrically shaped domains of illumination, utilised in our experiments, provided sharp boundaries between illuminated and non-illuminated areas. They did not form smooth gradients of illumination. Also size of the illumination domains was the same order as a `wavelength' of propagating plasmodium waves. These guaranteed that the illuminated shapes acted rather as reflectors, or mirrors,  than  sources of repelling fields. Therefore we envisage findings presented in the paper can be used in experimental implementation of collision-based computing schemes~\cite{adamatzky_cbc}, where plasmodium waves represent quanta of information and illumination domains are used to route momentary wires (trajectories) of the information quanta.

\end{document}